\documentclass[twocolumn,prb,showpacs,superscriptaddress]{revtex4}

\usepackage{graphicx}
\usepackage{dcolumn}
\usepackage{amsmath}
\usepackage{latexsym}
\usepackage{bm}

\newcommand{\ket}[1]{\left| {#1} \right>}
\newcommand{\bra}[1]{\left< {#1} \right|}

\makeatletter
\def\btt#1{\texttt{\@backslashchar#1}}%
\DeclareRobustCommand\bblash{\btt{\@backslashchar}}%
\makeatother

\begin{document}

\preprint{Paper}
\title{Errors due to Finite Rise/fall Times of Pulses 
                    in Superconducting Charge Qubits}
\author{Sangchul \surname{Oh}}
\email{scoh@mrm.kaist.ac.kr}
\affiliation{Department of Physics,
             Korea Advanced Institute of Science Technology, 
             Daejon 305-701, Korea}
\date{\today}
\begin{abstract}             
We study numerically the dynamics of two-qubit gates with 
superconducting charge qubits. The exact ratio
of $E_J$ to $E_L$ and the corresponding operation time
are calculated in order to implement two-qubit gates. 
We investigate the effect of finite rise/fall times of pulses 
in realization of two-qubit gates. It is found that the error 
in implementing two-qubit gates grows quadratically 
in rise/fall times of pulses.  
\end{abstract}
\maketitle
 
\section{Introduction}
\label{sec:lntro}
Building a practical quantum computer with a large number 
of qubits has recently attracted much attention since 
the useful quantum algorithms~\cite{Shor94,Grover97} 
and the quantum error-correction codes~\cite{Shor95,Steane96} 
were developed. Basic quantum logic gates were implemented on
trapped ions~\cite{Cirac95,Monroe95}, QED cavities~\cite{Turchette95}, 
and liquid state NMR~\cite{Chuang98,Cory98,Jones98}.
Solid-state devices have advantages of large scale integration, 
flexibility in the design, and easy connection to conventional 
electronic devices. Some of solid-state devices proposed 
are as follows: electron spins in quantum dots~\cite{Loss98}, 
nuclear spins of donor atom in silicon\cite{Kane98}, 
and ultrasmall Josephson junctions~\cite{Shnirman97,Averin98,Mooij99,
Makhlin99,Shnirman98,Makhlin01}.

Two types of superconducting qubits based on ultrasmall Josephson 
junctions were proposed. One is to use a number of excess Cooper 
pairs on a superconducting Cooper-pair box, called the 
superconducting charge qubit~\cite{Shnirman97,Averin98,Makhlin99}.
And the other is to utilize a single flux quantum of 
a superconducting loop, called the superconducting flux
qubit~\cite{Mooij99}. Here, we focus on superconducting charge 
qubits. In theoretical aspects, the measurement 
of superconducting charge states using a single electron 
transistor and decoherence due to coupling of qubits with 
environments were studied~\cite{Shnirman98,Makhlin01}. 
Also the quantum leakage of superconducting 
charge qubits was pointed out~\cite{Fazio99}. 
In experiments, Nakamura {\it et al.}~\cite{Nakamura99}
demonstrated the coherent oscillations of Cooper pairs 
on a superconducting Cooper-pair box. This corresponds to
a rotation of a single qubit about $x$ axis.  
Thus the realization of two-qubit gates is an important step for making 
a superconducting quantum computer with a medium size. 

In this paper, we investigate how the two-qubit gate 
could be implemented on superconducting charge qubits. 
The pulse sequence and operation times, and 
the simulation of two-qubit gates are reported.
We examine errors due to finite rise/fall times of pulses 
in the realization of two-qubit gates.

Our paper is organized as follows.
In Sec.~\ref{sec:Hamiltonian} we introduce the Hamiltonians 
of superconducting charge qubits and of ideal qubits. 
A numerical method which allows us to simulate a quantum 
computer is presented. In Sec.~\ref{sec:CN} we explicitly 
show the pulse sequence necessary to implement two-qubit 
gates with superconducting charge qubits. Also numerical 
studies of two-qubit gates are presented.
In Sec.~\ref{sec:error} we analyze errors due to finite 
rise/fall times of pulses on superconducting charge qubits. 
Finally, in Sec.~\ref{sec:summary} we summarize 
the results.

\section{Hamiltonian of superconducting charge qubits}
\label{sec:Hamiltonian}

In this paper we consider two qubit systems: ideal qubits
and superconducting charge qubits~\cite{Makhlin99,Makhlin01}. 
By comparing two qubit systems, one can discover differences 
and similarities between them, which will be helpful to improve 
the design of superconducting charge qubits. 

First consider an ideal model for a quantum computer.
The Hamiltonian of the system of $N$ ideal qubits reads
\begin{eqnarray}
 H_q^{\rm ideal}(t) 
     &=& -\sum_{i=1}^N {\bf B}_i(t)\cdot\bm{\sigma}^{(i)} 
                        \nonumber\\
     &{}&-\frac{1}{2} \sum_{i<j} J_{ij}(t)
           \bigl[\,\sigma_x^{(i)}\sigma_x^{(j)} + 
                   \sigma_y^{(i)}\sigma_y^{(j)}\,\bigr]\,,
\label{Hamil_ideal}
\end{eqnarray}
where $\bm{\sigma}^{(i)} = 
(\sigma_x^{(i)},\sigma_y^{(i)},\sigma_z^{(i)})$ 
are Pauli matrices for the $i$-th qubit. 
Here the Zeeman coupling terms ${\bf B}_i(t)$ 
and the inter qubit couplings $J_{ij}$(t) can be turned 
on and off between zero and finite values in a controlled way.

Let us consider a system of $N$ superconducting charge qubits
~\cite{Shnirman97,Makhlin99,Makhlin01}. 
Each qubit consists of a single-Cooper-pair box with two 
ultrasmall Josephson junctions of capacitance $C_J^0$ 
forming a DC-SQUID ring and a gate electrode with capacitance $C_g$. 
The dynamics of a qubit is characterized by relevant energy 
scales; superconducting gap $\Delta$, 
charging energy $E_C\equiv e^2/2(C_g + 2C_J^0)$, 
Josephson coupling energy $E_J^0$, and thermal fluctuation 
$k_BT$. Assume that the system is in
the regime of $E_C\ll \Delta$ and $k_BT\ll E_C$ in order 
to suppress quasi-particle tunneling or excitation.
Also suppose that the system operates under the conditions 
$E_J \ll E_C$ and $C_gV_g/(2e) \sim 1$. Then only two charge 
states $\{\ket{0},\ket{1}\}$, no excess Cooper pair and 
one excess Cooper pair on the box, play a role and 
represent the qubit. A few ways of coupling charge 
qubits were proposed~\cite{Shnirman97,Averin98,
Makhlin99,Makhlin01}. Here we consider the coupling
between charge qubits via the {\it LC} resonant circuit, 
where $N$ charge qubits are connected in parallel to 
a common inductor with inductance $L$. 
The Hamiltonian of the system of $N$ superconducting 
charge qubits is given by
\begin{eqnarray}
H_q(t) &=& -\frac{1}{2}\sum_{i=1}^N
           \bigl[\, E_{Ci}\sigma_z^{(i)} 
             + E_{Ji}(\Phi_{Xi})\sigma_x^{(i)} 
           \,\bigr] \nonumber\\
       && -\sum_{i\ne j}\frac{E_{Ji} E_{Jj}}{E_L} 
           \sigma_y^{(i)}\sigma_y^{(j)}\,,
\label{Hamil_squbit}
\end{eqnarray}
where $E_{Ci}=4E_C(C_{gi}V_{gi}/e - 1)$ is turn on 
and off by applying the gate voltage $V_{gi}$ on the 
$i$-th gate electrode. Here $E_{Ji}(\Phi_{Xi}) = 
2E_J^0\cos(\pi\Phi_{Xi}/\Phi_0)$ is the effective 
Josephson energy of the $i$-th qubit, and controlled 
by the external flux $\Phi_{Xi}$. $\Phi_0 = \hbar/2e$
is a flux quantum. 
Here is $E_L = \frac{\Phi_0^2}{\pi^2L}
\bigl(\frac{2C_J^0}{C_{\rm qb}}\bigr)^2$
with $C_{\rm qb}^{-1} = (2C_J^0)^{-1} + C_g^{-1}$.

The Hamiltonian of Eq.~(\ref{Hamil_squbit}) is similar 
to the ideal one of Eq.~(\ref{Hamil_ideal}) except one fact. 
The system of superconducting charge qubits 
has only two kinds of independently controllable parameters, 
$\{ E_{Ci}, E_{Ji}\}$, whereas the ideal model has four 
kinds of controllable parameters $\{B_{xi}, B_{yi}, B_{zi}, J_{ij}\}$. 
Coupling between two superconducting charge qubits could be
realized by turning on both $E_{Ji}$ and $E_{Jj}$. 
This causes somewhat disadvantages in operating 
two-qubit gates, which will be discussed in Sec.~\ref{sec:error}.

The time-evolution of $N$ qubits is governed by 
a time-dependent Schr\"odinger equation
\begin{eqnarray}
\label{Schroedinger}
i\hbar\frac{\partial}{\partial t} \ket{\psi} 
= H_q(t)\ket{\psi}\,,
\end{eqnarray}
where $\ket{\psi} = \sum_{m=0}^L a_m(t) \ket{m}$
with $L\equiv 2^N-1$ and $a_m \in {\bf C}$. 
Its computational basis is represented by a tensor product of 
individual qubits, 
$\ket{m} = \ket{q_1}\otimes\ket{q_2}\otimes\cdots\otimes\ket{q_N}$,
where integer $m = 2^{N-1}q_1 + 2^{N-2}q_2 + \cdots + 2^0 q_N$
and $q_i \in \{0,1\}$. Multiplying both sides of
Eq.~(\ref{Schroedinger}) by $\bra{n}$, one gets a set of 
$N$ coupled first-order ordinary differential equations 
for the function $a_n(t)$ 
\begin{eqnarray}
 \dot{a}_n = - \frac{i}{\hbar}\, 
               \sum_{m=0}^{L} H_{nm}(t)\, a_m \,,
\end{eqnarray}
where $H_{nm}(t) \equiv \bra{n} H_q(t) \ket{m}$.
We developed a program which allows us to solve the above initial 
value problem and to simulate a quantum computer. 
Our code is based on the Runge-Kutta method~\cite{press92}.

\begin{figure}[tbp]
\includegraphics{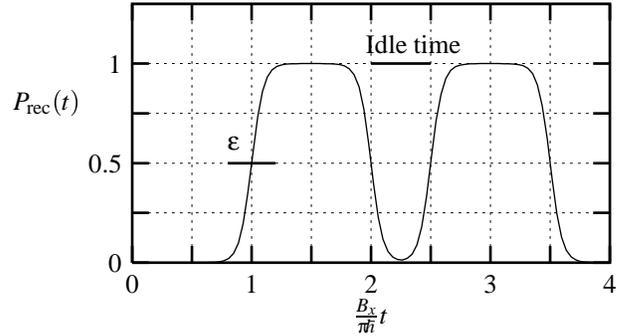}
\caption{ Rectangular pulses with finite rise/fall time $2\epsilon$ 
          are plotted as a function of time. Here time is normalized 
          in the unit of $\pi\hbar/B_x$. The role of idle time 
          between two successive pulses is to prevent the tails 
          of pulses from overlapping each others.}
\label{fig:rect-pulse}
\end{figure}

A quantum logic gate can be realized by controlling 
the time-evolution operator which acts on selected 
qubits for a fixed period of time. A way to control 
the time-evolution operator is to turn on or off each terms 
in the Hamiltonians of Eq.~(\ref{Hamil_ideal}) 
or~(\ref{Hamil_squbit}). For an example, a pulse $P(t)$, 
which is applied as $B_{\alpha i}(t) = B_{\alpha i}^0P(t)$ 
with $\alpha = x,y,z$, makes it possible to change
a term of the Hamiltonian from zero to finite value 
$B_{\alpha i}^0$ for a finite time $\tau$.
In this paper, we consider a rectangular pulse $P_{\rm rec}(t)$ 
with width $\tau = t_b - t_a$ and a unit height, 
modeled by a kink and anti-kink pair
\begin{eqnarray}
P_{\rm rec}(t) = 
        \frac{1}{2}\biggl[ \tanh\biggl(\frac{t- t_a}{\epsilon/2}\biggr)
      + \tanh\biggl(\frac{t_b - t}{\epsilon/2} \biggr)\biggr]\,,
\end{eqnarray}
where the rise (or fall) time is about $2\epsilon$. 
That means $P_{\rm rec}(t_a+\epsilon) \gtrsim \tanh(2)\approx
0.964$ for $\tau > 2\epsilon$. 
The smaller $\epsilon$ one takes, the sharper rectangular pulse one obtains.
Fig.~\ref{fig:rect-pulse} illustrates a sequence of two rectangular 
pulses. 

\section{CNOT gate with superconducting charge qubits}
\label{sec:CN}

In this section, we explicitly show how the controlled-not 
(CNOT) gate with superconducting charge qubits could be 
implemented by applying a sequence of pulses. 
What makes the CNOT gate so important is that the CNOT gate 
(or any nontrivial two-qubit gate) and single-qubit gates 
form a universal set of logic operation. Also a general 
two-qubit controlled-$U$ gate can be built up 
of two CNOT gates and three single-qubit gates. 
The CNOT gate acting on two qubits $i$ and $j$ is represented 
by the unitary matrix in the basis of 
$\{\ket{0_i0_j},\ket{0_i1_j},\ket{1_i0_j},\ket{1_i1_j}\}$ 
\begin{eqnarray}
U_{\rm CNOT}^{ij} = \left(\begin{array}{cccc}
                     1  &0  &0 & 0 \\
                     0  &1  &0 & 0 \\
                     0  &0  &0 & 1 \\
                     0  &0  &1 & 0 
                    \end{array}\right) \,,
\label{eq:CNOT}
\end{eqnarray}
where $i$ and $j$ are indices for a control bit and 
a target bit, respectively.

\subsection{CNOT gate with ideal qubits}
\label{subsec:CN-ideal}

First consider the implementation of the CNOT gate for 
an ideal system whose Hamiltonian is given by Eq.~(\ref{Hamil_ideal}).
The primary two-qubit gate acting on the $i$-th and $j$-th qubits 
could be implemented by turning on the coupling $J_{ij}(t)$. 
It can be written in the basis 
$\{ \ket{0_i0_j},\ket{0_i1_j}\ket{1_i0_j}, \ket{1_i1_j} \}$ as
\begin{eqnarray}
U_{\rm 2b}^{ij}(\gamma) 
= \left(\begin{array}{cccc}
   1  &0           &0           & 0 \\
   0  &\cos\gamma  &i\sin\gamma & 0 \\
   0  &i\sin\gamma &\cos\gamma  & 0 \\
   0  &0           &0           & 1 
  \end{array}\right) \,,
\end{eqnarray}
where $\gamma\equiv J_{ij}t/\hbar$. Then, the CNOT gate 
for an ideal qubit system can be realized by combining 
single-qubit gates and the primary two-qubit gate $U_{\rm 2b}^{ij}$
\begin{eqnarray}
\label{eq:CN-ideal}
U_{\rm CNOT}^{ij}  
          &=& {\rm  H}_i\, e^{i\frac{\pi}{4}}\,
              e^{-i\frac{\pi}{4}\sigma^{(j)}_x}\,
              e^{i\frac{\pi}{4}\sigma^{(i)}_x} \nonumber\\
          &&\times    U_{\rm 2b}^{ij}(\textstyle{\frac{\pi}{4}})\, 
              e^{ i\frac{\pi}{2}\sigma^{(i)}_x}\, 
              U_{\rm 2b}^{ij}(\textstyle{\frac{\pi}{4}})\,{\rm H}_i\,,
\end{eqnarray}
where ${\rm H}_i$ is a Hadamard gate acting on the $i$-th qubit.

\begin{figure}
\includegraphics{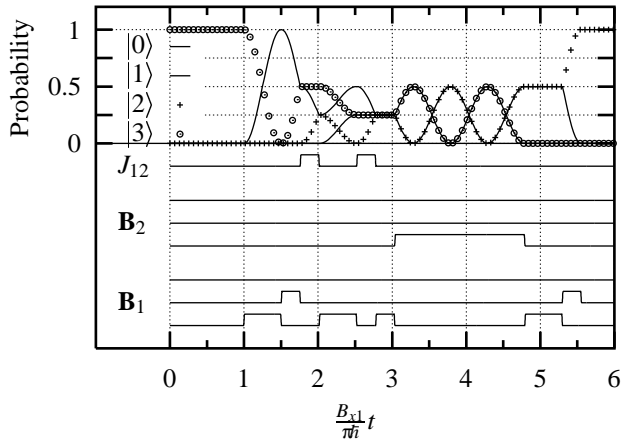}
\caption{Time-dependence of probabilities $|a_{ij}|^2$'s 
         as a function of the normalized time $B_{x1}t/\pi\hbar$
         on the action of the CNOT gate $U_{\text{CNOT}}^{12}$ in 
         an ideal model. The sequence of rectangular pulses is 
         given in the below part.
         Here is ${\bf B}_i=(B_{xi},B_{yi},B_{zi})$ with $i=1,2$.}
\label{fig:ideal-cnot}
\end{figure}

Fig.~\ref{fig:ideal-cnot} depicts the time-evolution of qubits
under the action of CNOT gate for the ideal model. The lower part of 
Fig.~(\ref{fig:ideal-cnot}) illustrates the sequence of rectangular 
pulses which generates the CNOT gate of Eq. (\ref{eq:CN-ideal}).
Here is taken $\ket{\psi_{\rm in}} = \ket{3} = \ket{11}$, i.e., 
$|a_{11}|^2 = 1$, as an initial state (denoted by $\circ$). 
After implementing the CNOT gate, one obtains $|a_{10}|^2 = 1$, 
corresponding to state $\ket{10}$, labeled by $+$. 
In Fig.~\ref{fig:ideal-cnot}, it takes a long time to implement 
the single-qubit gate $e^{-i\frac{\pi}{4}\sigma^{(j)}_x}$ because
$B_{\alpha i}$'s are taken to be always positive or zero.

\subsection{CNOT gate with superconducting charge qubits}
\label{subsec:CN-squbit}

Let us consider the implementation of the CNOT gate with 
superconducting charge qubits. The coupling between the $i$-th and 
$j$-th qubits can be switched on by turning on Josephson couplings, 
$E_{Ji}$ and $E_{Jj}$, and by turning off charging energy terms, 
$E_{Ci} = E_{Cj} = 0$. Then, the interaction 
Hamiltonian between two charge qubits, for an example qubits 1 and 2, 
can be written as
\begin{eqnarray}
H_{\text{ph}} &=& -\frac{E_J}{2}\sigma_{x}^{(1)} -\frac{E_J}{2}\sigma_{x}^{(2)}
      -E_{\rm int}\sigma_{y}^{(1)}\sigma_{y}^{(2)} \,,
\label{Hamil_ph}
\end{eqnarray}
where $E_{\rm int}\equiv {E_J^2}/{E_L}$ with an assumption of 
$E_{J1} = E_{J2} \equiv E_J$. For a moment, suppose that switching 
$E_{Ji}$'s on and off could be done instantaneously. As will be 
discussed later, a finite time in switching on $E_{Ji}$'s gives 
rise to an error in implementing two-qubit gates.

The basic two-qubit gate is given by the 
time-evolution operator under the Hamiltonian of Eq.~(\ref{Hamil_ph}) 
for a finite time $\tau$
\begin{eqnarray}
U_{\rm ph}(\tau) = e^{-iH_{\rm ph}\tau/\hbar}\,.
\end{eqnarray}
Let us transform the above time-evolution operator under 
the unitary operator $R_y \equiv 
\exp\bigl[-i\frac{\pi}{4}(\sigma_{y}^{(1)} + \sigma_{y}^{(2)})\bigr]$
as
\begin{eqnarray}
U_{\rm ph}' \equiv
R_y^\dagger\, U_{\rm ph}\, R_y = e^{-iH_{\rm ph}'\tau/\hbar}\,,
\end{eqnarray}
where the Hamiltonian transformed, 
$H_{\rm ph}' \equiv R_y^\dagger H_{\rm ph}  R_y$
reads
\begin{subequations}
\label{Hamil_tr}
\begin{eqnarray}
H_{\rm ph}'&=& -\frac{E_J}{2}\sigma_{z}^{(1)} -\frac{E_J}{2}\sigma_{z}^{(2)}
        -E_{\rm int}\sigma_{y}^{(1)}\sigma_{y}^{(2)}\\
&=& -E_{\rm int}
    \begin{pmatrix}
         a & 0 & 0 & -1 \\
         0 & 0 & 1 &  0 \\
         0 & 1 & 0 &  0 \\
        -1 & 0 & 0 & -a
    \end{pmatrix}\,.
\end{eqnarray}
\end{subequations}
Here $a\equiv {E_J}/{E_{\rm int}} = {E_L}/{E_J}$. 
Note that the Hamiltonian of Eq.~(\ref{Hamil_tr}) does not 
mix the subspace spanned by $\ket{00}$ and $\ket{11}$ 
with that by $\ket{01}$ and $\ket{10}$.
Then the time-evolution operator of 
the transformed Hamiltonian $H_{\rm ph}'$ 
becomes
\begin{eqnarray}
U_{\rm ph}'= \hspace{6.5cm}\nonumber\\
\begin{pmatrix}
\cos\theta + n_z\sin\theta &0         &0         &-in_x\sin\theta \\
       0                   &\cos\phi  &i\sin\phi &0            \\
       0                   &i\sin\phi &\cos\phi  &0            \\
     -in_x\sin\theta       &0         &0         &\cos\theta- n_z\sin\theta\\
\end{pmatrix},
\end{eqnarray}
where the time evolution corresponds to the rotation about
the $y$-axis by angle $\phi\equiv E_{\rm int}\tau/\hbar$
in the subspace spanned by $\ket{01}$ and $\ket{10}$.
Also in the subspace spanned by $\ket{00}$ and $\ket{11}$, 
the time evolution gives rise to the rotation about the axis $(n_x,0,n_z)$
by angle $\theta \equiv \sqrt{1+a^2}\, E_{\rm int}\tau/\hbar$, where
\begin{subequations}
\begin{eqnarray}
n_z   &\equiv& \frac{a}{\sqrt{a^2 +1}}
             = \frac{E_L}{\sqrt{E_L^2 + E_J^2}} \,,\\
n_x   &\equiv& \frac{1}{\sqrt{a^2 +1}}
            = \frac{E_J}{\sqrt{E_L^2 + E_J^2}} \,.
\end{eqnarray}
\end{subequations}

By choosing the appropriate values $E_L$, $E_J$, and $\tau$,  
one can control the rotation angles $\phi$ and $\theta$.
Let us consider special angles $\phi = \frac{\pi}{4}(2m - 1)$ with 
$m = 1, 2,\cdots$, and $\theta = n\pi$ with 
$n = 0, \pm 1, \pm 2,\cdots$. Note that the case of even $n$ is same 
to that of odd $n$ up to the global phase $e^{i\pi}$. 
These angles can be obtained by taking the evolution-time 
$\tau= \frac{\pi}{4}(2m-1)\frac{\hbar}{E_{\rm int}}$, which is 
given by the relation of angle $\phi$. Then one has  
$\theta = \frac{\pi}{4}(2m-1)\sqrt{1+ a^2} = \frac{\pi}{4}
\textstyle{\sqrt{1 + E_L^2/E_J^2}} = n\pi$. 
Thus we obtain the relation between $E_L$ and $E_J$
\begin{eqnarray}
\frac{E_L}{E_J} = \sqrt{\biggl(\frac{4n}{2m-1}\biggr)^2 -1}\,,
\end{eqnarray}
where it could be accomplished by tunning $E_J$ or $E_L$. 
Note that one can not take $n=0$.
Also the evolution-time is given by
\begin{eqnarray}
\tau = \frac{\pi\hbar}{4E_J}\sqrt{(4n)^2 -(2m-1)^2}\,.
\end{eqnarray}

We have the basic two-qubit gate 
$U_{\rm ph} = R_y\,U_{\rm ph}'\,R_y^\dagger$ 
written by
\begin{eqnarray}
U_{\rm ph}
= \frac{1}{2}
    \begin{pmatrix}
    1 + e^{i\phi} &  0             &  0            & 1 - e^{i\phi} \\
    0            & 1 + e^{-i\phi} & 1 -e^{-i\phi} & 0             \\
    0            & 1 - e^{-i\phi} & 1+ e^{-i\phi} & 0             \\
    1 - e^{i\phi} & 0              &  0            & 1 + e^{i\phi} \\
    \end{pmatrix}\,,
\end{eqnarray}
where $\phi = \frac{\pi}{4}(2m - 1)$ with $m = 1, 2,\cdots$. 
By combining $U_{\rm ph}$ with single-qubit gates,
the controlled phase flip gate $U_{\rm CPF}^{ij}$,
operating on the $i$-th and $j$-th qubits, can be 
realized by
\begin{eqnarray}
U_{\rm CPF}^{ij} &=& e^{-i\phi\sigma_z^{(i)}}\, 
                     e^{i\phi\sigma_z^{(j)}}\,
                     U_{\rm ph}^{ij}\, e^{-i\pi\sigma_z^{(i)}/2}\,
                     U_{\rm ph}^{ij} \,,
\end{eqnarray}
where $\phi$ is the value given above.
By using the controlled phase flip gate $U_{\rm CPF}^{ij}$ and 
the Hadamard gate H$_j$ on qubit $j$, the CNOT gate can be 
implemented by
\begin{eqnarray}
U_{\rm CNOT}^{ij} &=& {\rm H}_j U_{\rm CPF}^{ij} {\rm H}_j \,.
\end{eqnarray}

\begin{figure}
\includegraphics{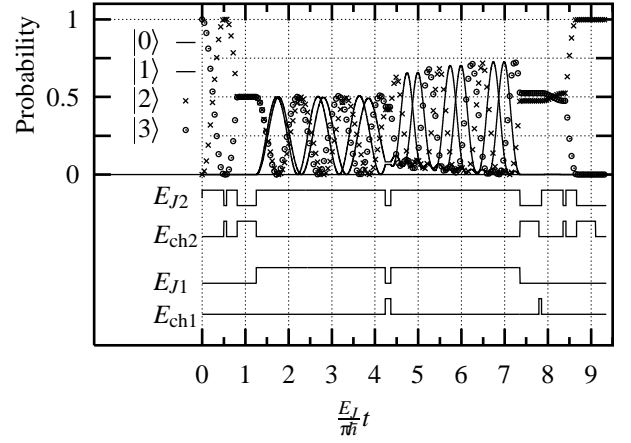}
\caption{ Time-evolution of probabilities of qubits as 
          a function of time on action of the CNOT gate for 
          superconducting charge qubits. 
          Time is normalized in the unit of $\pi\hbar/E_J$.
          $E_{Ci}/E_{J}= 2$ and $E_L = \sqrt{143}E_J$ are 
          taken.
          The sequence of rectangular pulses is
          depicted in the below.}
\label{fig:cnot_super}
\end{figure}

Taking $m = 1$ and $n=3$, one gets $E_L = \sqrt{143}\,E_J$.
This value satisfies the physical condition $E_L \sim 10 
E_J$~\cite{Makhlin99,Makhlin01,Dreher99}. 
The typical time-scale for operating the basic two-qubit gate 
is given by $\tau = \frac{\pi\hbar}{4E_J}\sqrt{143}$, 
which is much longer 
than the operation time of the single-qubit gate, 
$\tau_{\rm op} = \hbar/E_J$ or $\hbar/E_{Ci}$.
Fig.~\ref{fig:cnot_super} shows time-evolution of qubits when 
$\ket{\psi_{\rm in}}=\ket{11}$ is taken as an input 
(denoted by $\circ$). After operating the CNOT gate on two qubits, 
one gets the output $\ket{\psi_{\rm out}}=\ket{10}$, which 
is labeled by $\times$.

\section{Errors due to finite rise/fall times of pulses}
\label{sec:error}
 
Up to now, it was assumed that the effective Josephson coupling 
$E_{Ji}(t)$ could be switched on (or off) instantaneously. 
This means a pulse applied is a perfect rectangular one and has no 
rise/fall times. However, in reality it takes finite times 
to turn on (or off) pulses fully. While $E_{Ji}(t)$'s being 
switched on (or off), the Hamiltonian of Eq.~(\ref{Hamil_ph})
$H_{\rm ph}$ becomes time-dependent and the $H_{\rm ph}$'s 
at different times do not commute, i.e., 
$[H_{\rm ph}(t),H_{\rm ph}(t')] \ne 0$ for $t \ne t'$. 
This causes an error in the realization of two-qubit gates with 
superconducting charge qubits. If Hamiltonian's at different 
times commute, the shape of the pulse applied is of no importance. 
The ideal qubits and single-qubit rotations of superconducting 
charge qubits are free from this problem.

The Magnus expansion~\cite{Slichter92} provides the means of 
representing the time-evolution operator of a time-dependent 
Hamiltonian $H(t)$ as 
\begin{subequations}
\begin{eqnarray}
U(t) = T\,e^{-\frac{i}{\hbar}\int_{0}^{\tau}\,d\tau\,H(\tau)} 
     = e^{-\frac{i}{\hbar}\bar H\tau} \,,
\end{eqnarray}
where $T$ denotes the time-ordering operator.
Here the average Hamiltonian is given by
$\bar{H} = \bar{H}^{(0)} + \bar{H}^{(1)} + \cdots$
where
\begin{eqnarray}
\bar{H}^{(0)} &=& \frac{1}{\tau}\int_{0}^{\tau}\,dt_1\, H(t_1) \,,\\
\bar{H}^{(1)} &=& \frac{-i}{2\tau\hbar}\int_{0}^{\tau}\,dt_2\int_{0}^{t_2}\,dt_1
                  [H(t_2),H(t_1)]\,.
\end{eqnarray}
\end{subequations}
If the rise/fall time is short, then the approximation to
the first term $\bar{H}^{(0)}$ is good. Otherwise the higher order 
terms become significant.
If the pulse for the rise time is modeled by $P(t) = t/2\epsilon$,
the first term $\bar{H}^{(1)}$, which arises when the basic two-qubit 
gate $U_{\text{ph}}$ is implemented, can be written by
\begin{eqnarray}
\bar{H}^{(1)} \approx -\frac{E_J^3\epsilon}{15E_L\hbar}
                \Bigl(\sigma_z^{(1)}\sigma_y^{(2)}
                +\sigma_y^{(1)}\sigma_z^{(2)}\Bigr) \,.
\label{Hamil:H1}
\end{eqnarray}

The error due to the finite rise/fall times of pulses 
is quantified by the gate fidelity~\cite{Poyatos97}
\begin{eqnarray}
F=\langle\psi_{\rm in}| U^\dagger \rho_{\rm out} U
            |\psi_{\rm in}\rangle
 =|\langle\psi_{\rm out}|{\psi_{\rm out}^\epsilon}\rangle|^2\,,
\end{eqnarray}
where $U$ is the unitary operator corresponding to the ideal gate  
when perfect rectangular pulses are applied. On the other hand, 
the unitary operator generated by pulses with finite rise/fall times 
transforms the input state $\ket{\psi_{\text{in}}}$ into the density 
operator of the imperfect output state $\rho_{\rm out}$. 
In our case the gate fidelity is 
nothing but the square of the overlap between the perfect output state 
$\ket{\psi_{\text{out}}} = U\ket{\psi_{\text{in}}}$ 
and the imperfect output state  
$\rho_{\text{out}}=\ket{\psi_{\text{out}}^\epsilon}
\bra{\psi_{\text{out}}^\epsilon}$.
Assuming that Eq.~(\ref{Hamil:H1}) is small, it is straightforward
to calculate the gate fidelity 
\begin{equation}
F \approx|\langle\psi_{\text{in}}|e^{-i\bar{H}^{(1)}\tau/\hbar}
    \ket{\psi_{\text{in}}}|^2
  \approx 1 - \epsilon^2 \langle\Delta\eta^2\rangle
\label{Eq:fidelity}
\end{equation}
where $\langle\Delta\eta^2 \rangle \equiv
\bra{\psi_{\text{in}}}\eta^2\ket{\psi_{\text{in}}}
-(\bra{\psi_{\text{in}}} \eta\ket{\psi_{\text{in}}})^2$ is the 
dispersion of $\eta\equiv \bar{H}^{(1)}\tau/\hbar\epsilon$.
For small rise/fall time, error grows quadratically in rise/fall time 
$2\epsilon$. 

\begin{figure}[htbp]
\includegraphics{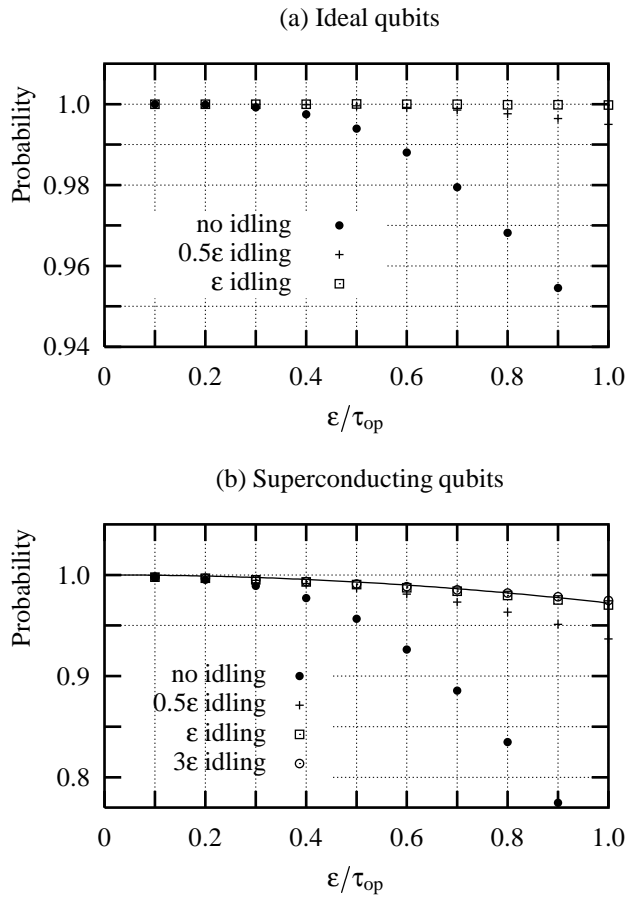}
\caption{Probability of success on the action of the CNOT gate
         as a function of the ratio of the rise/fall time
         to the operation time, $\epsilon/\tau_{\rm op}$: (a) 
         ideal qubits (b) superconducting charge qubits. 
         Here the time scale of operation is
         $\tau_{\rm op} = \hbar/B_z$ for the ideal case and
         $\tau_{\rm op} = \hbar/E_{Ci}$ for the superconducting 
         case. Idle times prevent the tails of pulses 
         from overlapping. The solid line is the fitting function 
         $1 - 0.027(\epsilon/\tau_{\rm op})^2$ which shows the quadratic
         growth of error in rise/fall times.}
\label{fig:switching-time}
\end{figure}

Fig.~\ref{fig:switching-time} shows the numerical study of
the error due to finite rise/fall times in implementing the CNOT gate 
with superconducting charge qubits. For ideal qubits, 
if finite idle times between successive pulses are applied, 
the correct CNOT gate is realized. However, 
for superconducting charge qubits, although a finite idle time 
reduces the error, there still exits the error due to finite 
rise/fall times. As shown in Fig.~\ref{fig:switching-time} (b),
the probability of getting the correct state reduces 
quadratically in rise/fall time $2\epsilon$ in agreement 
with the theoretical prediction of Eq.~(\ref{Eq:fidelity}).

In Nakamura {\it et al.}'s experiment~\cite{Nakamura99}, 
the rise/fall times of the pulse was about $30-40$ ps at 
the top of the cryostat. If the effective Josephson coupling 
is $E_J \approx 50$ $\mu$eV, then the timescale of single-qubit 
$x$ rotation is about $\hbar/E_J \approx 1$ ps. 
This means that the rise/fall times of the pulses should be less 
than 1 ps in order for two-qubit gates to be implemented correctly.
It should be noted that this type of errors is caused 
by the coupling scheme between two qubits. Thus this problems 
can be solved by improving the design of devices, i.e., 
by introducing new ways of two-qubit couplings.

\section{Summary}
\label{sec:summary}

We have studied the dynamics of qubits on 
the action of CNOT gates for ideal and superconducting charge qubits.
We have explicitly shown how the CNOT gate could be implemented for 
superconducting charge qubits. It is found that the error in implementing 
two-qubit gates with superconducting charge qubits 
grows quadratically in finite rise/fall times. 
Thus it is necessary to keep the rise/fall times small or 
to find new ways of coupling two qubits other than 
the coupling scheme via the common inductor. 

\section*{Acknowledgments}
The author would like to thanks Y. Makhlin, A. Shnirman, and
G. Sch\"on for helpful discussions. 
This work was partially supported by 
Korea Science and Engineering Foundation.

\end{document}